# Hidden Figures of Globular Clusters: Integrated Stellar Populations Impacted by Hot Subdwarfs

Thayse A. Pacheco[1,2], Paula R. T. Coelho[2], Lucimara P. Martins[3], Ricardo P. Schiavon[4], Erik V. R. de Lima[2], Marcos P. Diaz[2], Domenico Nardiello[5], Ronaldo S. Levenhagen[6], Rogério Riffel[1], Charles J. Bonatto[1], and Ana L. Chies-Santos[1]

[1] Universidade Federal do Rio Grande do Sul, Instituto de Física, Av. Bento Gonçalves, 9500, Porto Alegre, RS 91501-970, Brazil; thayse.pacheco@ufrgs.br, thayse.pacheco@gmail.com
[2] Universidade de São Paulo, Instituto de Astronomia, Geofísica e Ciências Atmosféricas, Rua do Matão, 1226 São Paulo, SP 05508-900, Brazil
[3] Universidade Cidade de São Paulo, Núcleo de Astrofísica, Rua Cesário Galero, 448/475 São Paulo, SP 03071-000, Brazil
[4] Liverpool John Moores University, Astrophysics Research Institute, 146 Brownlow Hill, Liverpool L3 5RF, UK
[5] Università degli Studi di Padova, Dipartimento di Fisica e Astronomia "G. Galilei," Via Francesco Marzolo, 8, 35121 Padova PD, Italy
[6] Universidade Federal de São Paulo, Departamento de Física, Rua Prof. Artur Riedel, 275, CEP 09972-270, Diadema, SP, Brazil

*Received 2025 June 9; revised 2025 August 22; accepted 2025 August 24; published 2025 October 10*

## Abstract

Globular clusters (GCs) are fundamental for understanding the integrated light of old stellar populations and galaxy assembly processes. However, the role of hot, evolved stars, such as horizontal branch (HB), extreme HB, and blue stragglers, remains poorly constrained. These stars are often underrepresented or entirely excluded from stellar population models, despite their dominant contribution to the ultraviolet (UV) flux. Their presence can bias age estimates by mimicking the spectral signatures of younger populations. We examined the impact of evolved hot stars on the models using two well-studied Galactic GCs with high-quality Hubble Space Telescope photometry and integrated spectra from the International Ultraviolet Explorer and the Blanco Telescope. NGC 2808 and NGC 7089 (M 2) have extended HBs and are proxies for old stellar populations. Integrated spectra were constructed using a color magnitude diagram–based (CMD-based) method, matching observed stars to evolutionary phases and then to appropriate synthetic stellar libraries, enabling the HB morphology to be taken into account. Our findings show that the inclusion of evolved hot stars significantly improves the agreement between the model and observed spectra from the UV to the optical. The inclusion of these phases reduced the residuals in spectral comparisons. Our results reinforce that comprehensive stellar population models incorporating evolved hot components are essential to accurately date unresolved systems and to robustly trace formation histories of extragalactic galaxies.

*Unified Astronomy Thesaurus concepts:* Stellar populations (1622); Horizontal branch (2048); Globular star clusters (656); Multi-color photometry (1077); Spectral energy distribution (2129); Subdwarf stars (2054)

## 1. Introduction

Globular clusters (GCs) are groups of stars born together, orbiting a common kinematical center under their collective gravitational potential. Due to their high mass and deep gravitational potential, they often survive for a Hubble time (J. P. Brodie & J. Strader 2006; M. A. Beasley 2020). GCs are formed mainly at the early beginning of the assembly of galaxies, representing relics of the first episodes of star formation in their host galaxies (J. Strader et al. 2005; A. L. Chies-Santos et al. 2011; M. A. Beasley 2020; K. A. Alamo-Martínez et al. 2021). They are key tracers of the assembly history and evolution of galaxies, including the Milky Way (D. A. Forbes & T. Bridges 2010). This research field is at the interface of Galactic archeology and extragalactic astrophysics, connecting these fields to stellar population studies.

The study of stellar populations can be divided into two primary approaches: resolved and integrated (or unresolved) light. Resolved stellar population studies involve analyzing individual stars within a specific region or as part of a nearby stellar system and providing a detailed characterization of their physical and chemical properties (C. Gallart et al. 2005; E. Tolstoy et al. 2009). However, integrated stellar population studies consider light emission from all stars within an extended object where most individual stars cannot be resolved, offering a perspective of combined stellar content (J. Walcher et al. 2011; C. Conroy 2013).

The methods of stellar population synthesis have been evolving for decades (W. A. Baum 1959; J. Crampin & F. Hoyle 1961; A. Sandage 1961; H. L. Johnson 1966; H. Spinrad 1966; B. M. Tinsley 1968; S. M. Faber 1972; B. M. Tinsley 1972; B. M. Tinsley 1973; E. Bica & D. Alloin 1986; E. Bica 1988). In the following discussion, we outline the main methodologies that are frequently used.

*Evolutionary-based methodologies.* The isochrone synthesis revolutionized stellar population studies by combining stellar evolution, spectral libraries, and initial mass functions (IMFs) for comprehensive galaxy modeling (G. A. Bruzual 1983; E. Bica & D. Alloin 1986; S. Charlot & G. Bruzual 1991; G. Bruzual & S. Charlot 2003; T. C. Moura et al. 2019; C. Rennó et al. 2020). This methodology can consider factors such as dust attenuation, star formation history (SFH), chemical evolution, and simple stellar populations (SSPs), all of which impact the mass-to-light ratio, luminosity evolution, and spectral energy distributions (SEDs).

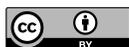

*The fuel consumption theorem.* The models of C. Maraston (2005), C. Maraston (2011), and C. Maraston & G. Strömbäck (2011) adopt the fuel consumption theorem proposed by A. Renzini & A. Buzzoni (1986). This method addresses the problem of discrete sampling in stellar evolution tracks, revealing the proportional relationship between the nuclear fuel burned in post–main sequence (Post–MS) evolutionary phases and its contribution to the total luminosity of the SSP. C. Maraston (1998) applied this theorem to illustrate the bolometric budget of SSP over time.

*CMD-based methodologies.* An alternative to isochrone-based synthesis and an ad hoc assumption of the IMF is to construct models directly from observed color–magnitude diagrams (CMD) and stellar spectral libraries. Each star in the CMD is matched to a representative spectrum from the library, and the integrated spectrum is built by summing the flux-weighted contributions of all stars in a given photometric band (J. F. C. J. Santos et al. 1990, 1995; R. P. Schiavon et al. 2002a, 2002b; L. P. Martins et al. 2019).

Stellar population synthesis based on observed CMDs avoids uncertainties in isochrones and is better suited to assess phases poorly predicted by stellar evolution, such as the horizontal branch (HB), extreme HB (EHB), and blue straggler (BS) phases. The loci of these phases in the HR diagram depend on assumptions about the mass loss at the tip of a red giant branch (A. V. Sweigart et al. 2002), binarity (I. Pelisoli et al. 2020), and He abundance.

The blue and ultraviolet (UV) flux in young GCs is dominated by the locus of the MS turnoff, while hot-evolved low-mass stars, such as HB, EHB, and BS stars, contribute most in older systems. Distinguishing young massive stars from old low-mass stars is crucial for dating extragalactic GCs, as age–HB morphology degeneracy complicates age estimates (L. Greggio & A. Renzini 1990).

Moderately metal-rich M31 GCs exhibit a mix of cool and hot HB stars (R. C. Peterson et al. 2003). EHB stars, such as hot subdwarfs, can be a source of the UV excess in old GCs and elliptical galaxies (S. Yi et al. 1999; G. Busso et al. 2005; E. M. Green et al. 2008). Likewise, the UV upturn in early-type galaxies is attributed to HB and EHB stars and their descendants, with S.-J. Yoon et al. (2004) suggesting it as an age indicator and S. K. Yi (2008) proposing a high He abundance as its origin. Therefore, to derive reliable ages, integrated stellar population models must account for HB and EHB stars.

Regarding the effect on spectral indices, J. A. de Freitas Pacheco & B. Barbuy (1995) showed that H$\beta$ index models must consider the morphology of HB along with the metallicity. This is supported by UV observations and suggests that a blue HB contributes to the H$\beta$ strength in an integrated stellar population (R. P. Schiavon et al. 2004b; R. P. Schiavon 2007; A. J. Cenarro et al. 2008). The H$\delta_F$/H$\beta$ ratio in old GCs is more sensitive to HB morphology than age, indicating differentiation between true intermediate age clusters and old ones with strong Balmer lines.

The age–HB morphology degeneracy also causes major uncertainty in spectral fitting age. P. Ocvirk (2010) found spuriously younger ages than the MS turnoff estimates and false bursts in the SFH of low-metallicity GCs, contributing up to 12% of the optical light. This problem persists in modern evolutionary-based models and widely used spectral fitting codes, causing age discrepancies of up to 50% (G. Gonçalves et al. 2020; R. Asa'd et al. 2025).

C. Maraston & D. Thomas (2000) combined old-metal-rich and old-metal-poor components in composite stellar population models, reproducing the observed H$\beta$ strengths in elliptical galaxies with purely old ages. The EHB was later incorporated into isochrone synthesis models by F. Hernández-Pérez & G. Bruzual (2013), considering binary interactions and He White Dwarf mergers to effectively explain the presence of EHB stars in metal-rich open clusters. I. Cabrera-Ziri & C. Conroy (2022) refined the spectral synthesis by including one hot HB star as a free extra component to the integrated light, improving the age agreement between the spectral fitting and the color–magnitude values for the Galactic GCs. Recently, I. Martín-Navarro & A. Vazdekis (2024) showed EHB features in spectral fitting, highlighting their impact on integrated light and the need to explicitly include such populations in synthesis to avoid biases in age and SFH estimates.

Hot, low-mass HB stars significantly affect optical light, mimicking younger ages, highlighting the importance of extending the analysis toward UV to resolve the age–HB degeneracy. Although most stellar population synthesis libraries are optimized for optical and near-IR wavelengths (e.g., the UVES Paranal Observatory Project, UVES-POP; S. Bagnulo et al. 2003; S. B. Borisov et al. 2023), the high-resolution spectral library dedicated to the UV, the UVBLUE library (L. H. Rodríguez-Merino et al. 2005), lacks high-gravity models. Therefore, improving spectral synthesis in the UV and far-UV, particularly for high-gravity models, is essential to predicting old stellar population ages more reliably.

Due to observational limitations, most stellar systems beyond the local Universe are only accessible through integrated observations (C. Conroy 2013). However, Galactic GCs can be observed in both resolved and integrated manners, making them crucial proxies for testing integrated-light models and methodologies. In this study, we employ a two-step strategy: first, a resolved study that enables a comprehensive mapping of individual stars, followed by using this information to model the integrated light emitted by the entire population.

In this work, we aimed to investigate the impact of hot-evolved (BS, HB, and EHB) stars brighter than the MS turnoff on the modeling of integrated spectra of old Galactic GCs using an adapted CMD-based methodology. We quantified the age differences derived from the spectral fitting of models with and without these hot-evolved components. Ultimately, we addressed the following question: How much do hot low-mass stars contribute to the integrated light of old stellar populations from the UV to the visible?

In Section 2, we provide an overview of two Galactic GCs that we use as a base population, NGC 2808 and NGC 7089 (M 2). In Section 3, we describe the photometric data obtained from Hubble Space Telescope (HST) observations. Section 4 introduces the stellar flux libraries and synthetic photometry employed in our analysis. Section 5 outlines our methodology, which includes tagging stars by their evolutionary phases and matching observed stars to synthetic spectra. Section 6 explains the construction of synthetic integrated spectra. Section 7 presents the effects of different stellar components in age determinations and comparisons with observational data. Finally, Section 8 summarizes our findings and conclusions.

## 2. Sample

We used old Galactic GCs as proxies for studying integrated, "simple" stellar populations. Despite these sources being composed of multiple stellar populations (see, e.g., N. Bastian & C. Lardo 2018), this is a first step toward accurately reproducing the integrated light under more realistic assumptions. Analyzing these Galactic GCs allows us to validate our methods regarding integrated stellar populations by comparing them with resolved observations. For this work, we selected two Galactic GCs, NGC 2808 and NGC 7089 (M 2), observed by the HST as featuring extended HBs.

### 2.1. NGC 2808

NGC 2808 is one of the oldest well-known Milky Way GCs, at $11.5 \pm 0.4$ Gyr old (G. Limberg et al. 2022), and its stars originated within 200 million year period. Also, as one of the most massive clusters, it includes more than 1 million stars and has a mass of $8.5 \times 10^5 \, M_\odot$ (D. E. McLaughlin & R. P. van der Marel 2005). There is evidence of up to five stellar populations and different abundances in NGC 2808. E. Carretta (2015) found a metallicity of [Fe/H] = $-1.129 \pm 0.005 \pm 0.034$ ($\pm$statistical $\pm$ systematic error). However, J. E. Colucci et al. (2017) showed differences between the mean abundances of the Fe I and Fe II lines, with final abundances of [Fe I/H] = $-1.04 \pm 0.04$ and [Fe II/H] = $-0.85 \pm 0.04$. A recent study by C. Lardo et al. (2022) confirms a spread of metallicity for several GCs, and C. Lardo et al. (2023) shows that NGC 2808 has a variation equal to $0.25 \pm 0.06$ dex. They also presented the mean abundance of [Fe/H] = $-1.03 \pm 0.07$ for the first population. Using high-resolution spectra from APOGEE DR17, G. Limberg et al. (2022) show a mean metallicity [Fe/H] = $-1.09 \pm 0.01$ with a dispersion of 0.05 dex, and this will be the reference value hereafter.

### 2.2. M 2

M 2 is one of the largest and oldest GCs in the Milky Way, $11.5 \pm 0.3$ Gyr old (G. Limberg et al. 2022). A. P. Milone et al. (2014) suggests up to seven stellar populations to explain its chemical evolution. The iron abundance distribution for M 2 presented by D. Yong et al. (2014) displays a prominent peak around [Fe/H] = $-1.7$, accompanied by smaller peaks at approximately $-1.5$ and $-1.0$, where the latter group was not established. However, C. Lardo et al. (2016) showed that the abundance of iron in M 2 is bimodal, with the first component at [Fe/H] = $-1.5$ and a small fraction of stars at [Fe/H] = $-1.1$. G. Limberg et al. (2022) shows a mean metallicity [Fe/H] = $-1.46 \pm 0.02$ with a dispersion of 0.05 dex, which will be used as the reference value in the following discussion.

## 3. Photometric Data

We used multiband photometric observations of Galactic GCs from HST (G. Piotto et al. 2015; D. Nardiello et al. 2018). The data cover a wide range of wavelengths, from UV to near-infrared (IR). It comprises the F275W, F336W, and F438W bands observed by the new Wide-field Camera 3 in the UV and visible light channel. It also includes the F606W and F814W bands observed by the Advanced Camera for Surveys in the Wide-field Channel. The stacked images for each GC with the exposure times given in each filter are publicly available in the HST UV Globular Cluster Survey (HUGS) archive (G. Piotto 2018).[7] Full details on the observation strategy, the combination of images, and the depth achieved can be found in Section 5 and Figure 5 of D. Nardiello et al. (2018), which documents the limits of the catalog coverage.

**Table 1**
Maximum Magnitudes in Each Filter

| Filter | NGC 2808 (mag) | NGC 7089 (mag) |
|---|---|---|
| F275W | 28.3 | 29.2 |
| F336W | 25.3 | 26.4 |
| F438W | 23.6 | 26.3 |
| F606W | 22.2 | 24.4 |
| F814W | 21.1 | 23.1 |

We selected stars that are likely members of the cluster, with a membership probability greater than 90% (D. Nardiello et al. 2018). We applied a threshold for the photometric errors, excluding observations with uncertainty exceeding 0.07 mag for the F275W, 0.05 mag for F336W, and 0.03 mag for the F438W, F606W, and F814W bands. The observed photometry was corrected for interstellar extinction by calculating the reddening values at each filter's effective wavelength. The extinction law of E. L. Fitzpatrick & D. Massa (2007), defined from 910 to 6000 Å, was employed using the Python package `extinction`, which has implemented several empirical laws of dust extinction (K. Barbary 2016). The photometric extinction was based on the color excess $E(B-V)$ of 0.22 and 0.06 for NGC 2808 and M 2, respectively (C. Usher et al. 2017). From this sample, we have listed in Table 1 the magnitude depth in each filter for both GCs after our selection and corrections.

## 4. Stellar Flux Libraries

In this study, we used three synthetic stellar spectral libraries, namely F. Castelli & R. L. Kurucz (2003), P. R. T. Coelho (2014), and T. A. Pacheco et al. (2021, 2023).

Spectra from T. A. Pacheco et al. (2021, 2023) were used to represent the EHB for both GCs, NGC 2808 and M 2. This is a grid of detailed stellar atmosphere models in NLTE (nonlocal thermodynamic equilibrium) for extremely blue HB stars, high-resolution NLTE spectra, and synthetic photometry, covering a temperature range of $10{,}000 \leqslant T_{\rm eff}$ [K] $\leqslant 65{,}000$, surface gravities of $4.5 \leqslant \log g$ [cgs] $\leqslant 6.5$, and two extreme scenarios of abundances of He (He rich and He poor). We also considered two sets with different metallicities: a solar metallicity case ([Fe/H] = 0) and one with halo stellar metallicity ([Fe/H] = $-1.5$, with an enhancement of [$\alpha$/Fe] = $+0.4$). In the following analysis, we used spectral models with chemical parameters that represent the typical Galactic halo with the He-rich sequence for all the studied GCs.

We chose models from the P. R. T. Coelho (2014) grid to represent the other evolutionary phases in NGC 2808. This work computed a theoretical stellar library that covers a temperature range of $3000 \leqslant T_{\rm eff}$[K] $\leqslant 25{,}000$ and includes solar and $\alpha$-enhanced compositions. It provides a wide range of atmospheric parameters, such as a surface gravity range of $-0.5 \leqslant \log g$[cgs] $\leqslant 5.5$ and 12 different chemical mixtures covering $0.0017 \leqslant Z \leqslant 0.049$ ($-1.3 \leqslant$ [Fe/H] $\leqslant +0.2$). These spectral models have a resolving power ($R$) of 20,000 and cover the wavelength range from 2500 to 9000 Å, with a constant sampling of 0.2 Å. From this set of parameters, we used the metallicity of [Fe/H] = −1.0 and enhancement of [$\alpha$/Fe] = +0.4.

---
[7] https://archive.stsci.edu/prepds/hugs/

**Table 2**
Synthetic Parameters Used in the Analysis

| Library | T. A. Pacheco et al. (2021, 2023) | P. R. T. Coelho (2014) | F. Castelli & R. L. Kurucz (2003) |
|---|---|---|---|
| # of models | 36 | 328 | 476 |
| $\lambda$ [Å] | 1000 to 10,000 | 2500 to 9000 | 1000 to 10,000 |
| $T_{\rm eff}$ [K] | 10,000 to 65,000 | 3000 to 26,000 | 3500 to 50,000 |
| $\log g$ [cgs] | 4.5 to 6.5 | −0.5 to 5.5 | 0.0 to 5.0 |
| s [Fe/H] | −1.5 | −1.0 | −1.5 |
| [$\alpha$/Fe] | +0.4 | +0.4 | +0.4 |
| $R_{\rm mean} = \frac{\lambda}{\Delta\lambda}$ | 250,000 | 20,000 | 250 |
| Sampling [Å] | 0.01 | 0.2 | 10 to 20 |

The model grid from F. Castelli & R. L. Kurucz (2003) was chosen to represent the evolutionary phases of M 2. We used the F. Castelli & R. L. Kurucz (2003) library instead of P. R. T. Coelho (2014), as the latter does not reach the low metallicity of this cluster. F. Castelli & R. L. Kurucz (2003) computed statistical surface fluxes that cover a temperature range of $3500 \leqslant T_{\rm eff}[{\rm K}] \leqslant 50,000$ and include solar and $\alpha$-enhanced compositions. This library provides a wide range of atmospheric parameters. These low-resolution SEDs mimic a resolving power ($R$) of approximately 250 and cover the wavelength range from 1 to 100,200 Å with nonconstant spectral sampling. From this set of parameters, we used wavelengths from 1000 Å to 10,000 Å, the metallicity of [Fe/H] = −1.5, and the enhancement of [$\alpha$/Fe] = +0.4. In the wavelength range of interest, the sampling comprises points every 10 Å at 1000 Å and 20 Å at 10,000 Å.

We selected only the metallicity closest to the target GC from each grid. Table 2 presents the number of models used in this work, with their respective range of wavelength $\lambda$, atmosphere parameters $T_{\rm eff}$ and $\log g$, chemical abundances [Fe/H] and [$\alpha$/Fe], R,[8] and wavelength sampling.

### 4.1. Synthetic Photometry

We adopted the VEGA system with respective zero-points as the reference flux value for each photometric band. For this section, we used the Spanish Virtual Observatory (SVO), which provides standardized and comprehensive access to astronomical data. It includes photometric band parameters, transmission curves, and calibration, facilitating multiwavelength studies (C. Rodrigo et al. 2012; C. Rodrigo & E. Solano 2020). The filters were interpolated in the spectra wavelength steps within the Shannon–Whittaker scheme (C. Shannon 1949). Photon-counting integrated fluxes were assumed (M. S. Bessell et al. 1998). Vega's zero-points were previously evaluated from Calspec's standard spectrum (R. C. Bohlin et al. 2014). Details about the identification associated with each filter, instrument information, effective, minimum, and maximum wavelengths, and their flux zero-points are presented on the SVO website.[9] Following interpolation, proper passband integration allows for flexible synthetic magnitude estimates from spectral library data. In addition, we computed the area under the curve of each synthetic spectrum, which was subsequently used in the spectral normalization process. We calculated the synthetic photometry of the three synthetic spectral grids for five different HST filters: F275W, F336W, F438W, F606W, and F814W.

## 5. Novel Methodology

The integrated spectra were synthesized using an adapted CMD-based process (J. F. C. J. Santos et al. 1990, 1995; R. P. Schiavon et al. 2002a, 2002b; L. P. Martins et al. 2019). This methodology avoids the caveats of isochrones that cannot explain evolved stellar populations, such as EHB, and one does not have to adopt an ad hoc IMF assumption. Here, a two-step variant of the method is presented, as described in the following discussion.

### 5.1. Tagging the Evolutionary Phase

The photometric data were categorized into six distinct groups for each GC, aiming to represent stellar evolutionary stages (see Figures 1 and 2):

1. MS: The stars are cooler and fainter than the visual turnoff.
2. Red and asymptotic giant branches (GB): The stars are cooler than the visual turnoff and brighter than the MS.
3. Red HB (RHB): The HB stars between the HB gap and the GB.
4. BS: The plume of stars brighter and hotter than the turnoff and fainter than the HB.
5. Blue HB (BHB): The intermediate between RHB and EHB.
6. Extreme HB (EHB): The stars hotter than the knee of the HB in the UV color (F275W–F336W).

These phases are classified based on their observational features in different filters. The selection process to create those subsamples was done using TOPCAT (Tool for OPerations on Catalogs And Tables; M. B. Taylor 2005). The cutoffs for each GC are shown in Tables 4 and 5 in Appendix A. Note that by applying the color cutoffs, some stars of the bottom MS are excluded from the sample selection presented in Section 3. This explains the lower magnitude depth in the CMDs compared to Table 1. While completeness decreases at the faint end of MS in the edges of the HST images, the stars that dominate the integrated light (GB, HB, and upper MS) are well represented. Artificial star tests, as described in Appendix B, support that this level of incompleteness does not significantly affect the integrated flux.

The resulting CMDs with the stars of NGC 2808 tagged by evolutionary phase are presented in Figure 1. Similarly, the CMDs for M 2 with stars tagged by evolutionary phase are shown in Figure 2.

Both GCs analyzed in this study exhibit complex enrichment histories (see Section 2). This level of complexity is not unusual in extragalactic systems, where a mix of multiple populations is common. Although such complexity may introduce additional scatter in derived stellar population parameters, it also highlights the strength of our method. Our CMD-based approach remains robust in the presence of multiple stellar populations, accounting for the full spread of stars in the CMD rather than fitting a single isochrone. This observed feature may reflect a range of metallicities and could be more accurately modeled in future work by employing a grid of stellar models with varying metallicities. This provides a more realistic strategy for modeling integrated spectra in systems with multiple stellar populations.

---
[8] At $\lambda = 5000$ Å.
[9] https://svo2.cab.inta-csic.es/theory/fps/

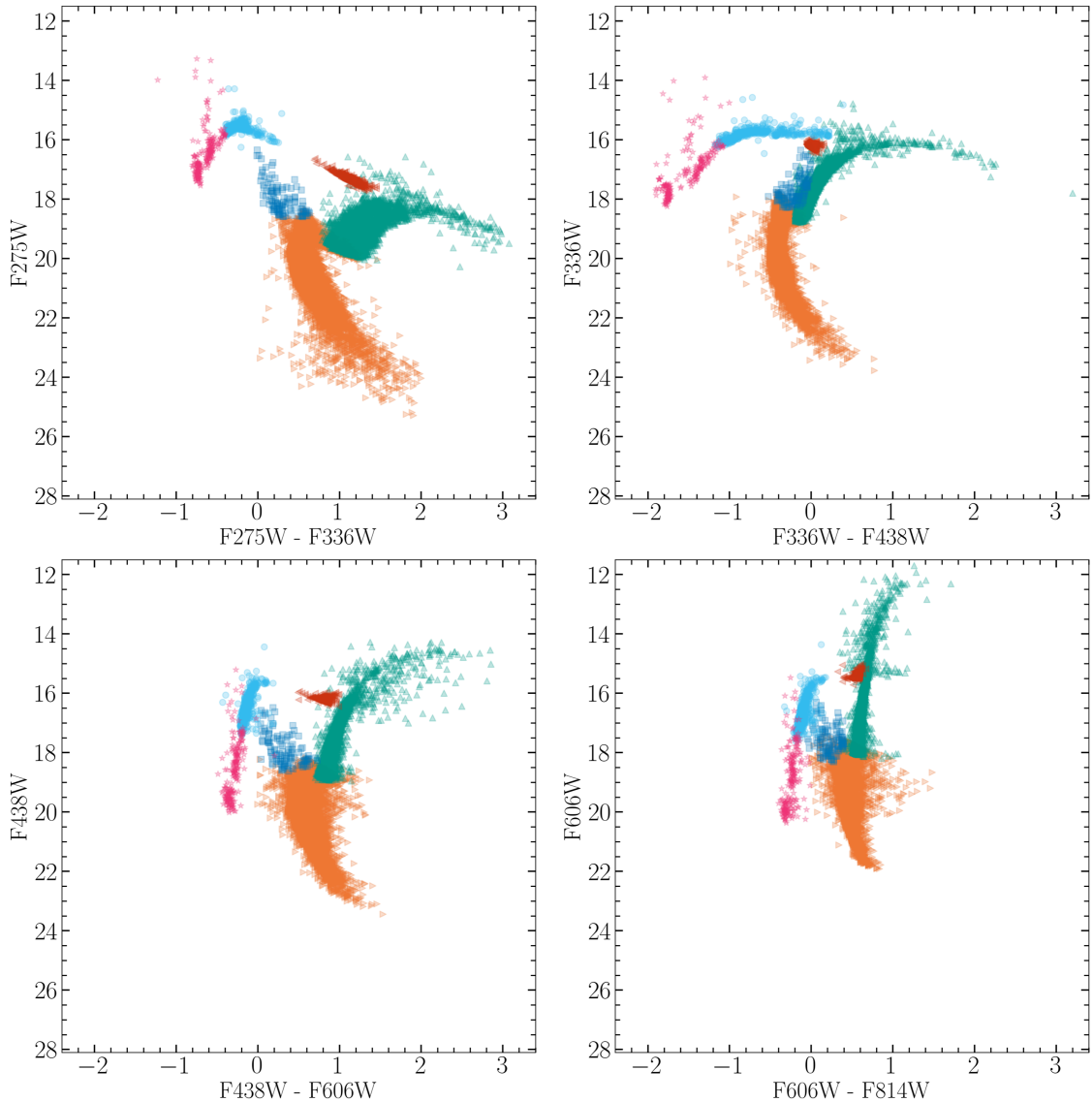

**Figure 1.** CMDs of NGC 2808, with stars color-coded by evolutionary subset: MS (orange right-facing triangles), GB (teal upward-facing triangles), RHB (red left-facing triangles), BS (blue squares), BHB (cyan points), and EHB (magenta stars). These show how CMDs in different bands are varied in morphology, particularly the BHB and EHB.

### 5.2. Matching Observed to Synthetic Stars

Usually, CMD-based methodologies use power-law transformations from the observational plane (color and magnitude) to the theoretical plane ($T_{\rm eff}$ and $\log g$). L. P. Martins et al. (2019) used the high-order fitting functions of the G. Worthey & H.-C. Lee (2011) calibration for transforming the data into the Kiel diagram. The primary source of error in color-temperature calibration is finding a suitable polynomial to trace the intricate variations in colors accurately. In addition, the fitting process involves a substantial number of stars for each fit. In particular, uncertainties for cool dwarf stars are more significant than uncertainties in giants due to the small number of dwarf calibrations to estimate it (G. Worthey & H.-C. Lee 2011). A significant limitation in this calibration is its applicability only to GB stars, which only covers some of the parameters required for our analysis.

We circumvent this translation by directly matching each observed star to a synthetic spectrum in a multidimensional color plane and avoiding the uncertainties associated with parametric transformations. In the method employed, both synthetic and observational photometry were combined in pairs of filters to build a 10-dimensional space: F275W−F336W, F275W−F438W, F275W−F606W, F275W−F814W, F336W−F438W, F336W−F606W, F336W−F814W, F438W−F606W, F438W−F814W, and F606W−F814W. This accounts for the variations in all 10 attributes, determining the overall separation between the two points in this higher-dimensional space. This method facilitates model selection with the closest representation of physical parameters for each star of a GC.

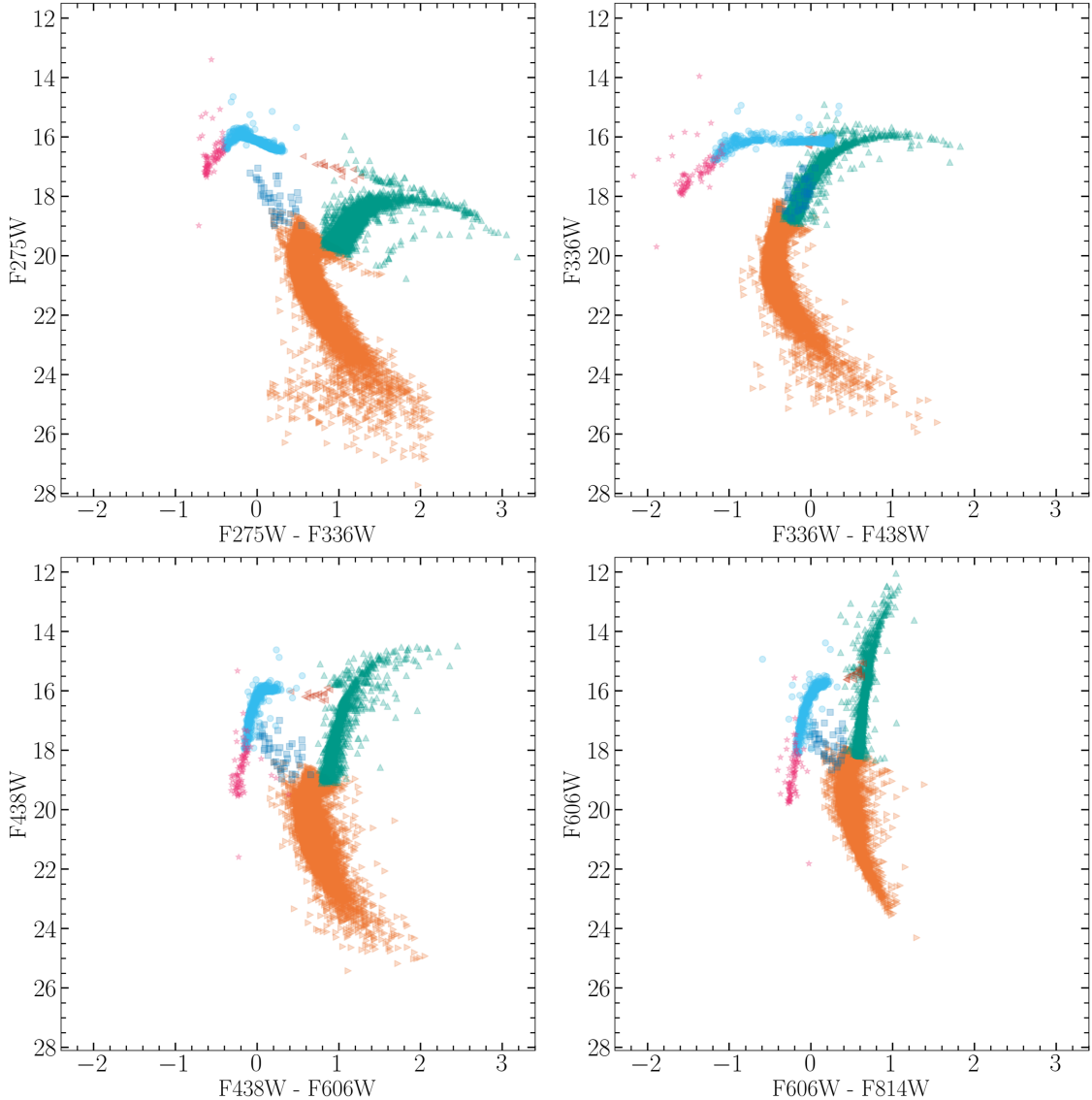

**Figure 2.** CMDs of M 2, with stars color-coded by evolutionary subset, following the same scheme as in Figure 1.

A model representing each star is selected by minimizing its color distance. We computed the geometrical distance between synthetic and observed colors in this 10-dimensional space. The closest model was located with the K-nearest neighbors (KNN) algorithm implemented in the `scikit-learn.neighbors` python package (F. Pedregosa et al. 2011). Figure 3 shows five color–color diagrams representing the 10 colors used in the analysis of the NGC 2808. The distribution of our sample of observed stars by evolutionary phases is hued in the same pattern as in Figure 1. The synthetic colors of P. R. T. Coelho (2014) are shown in gray crosses, and those of T. A. Pacheco et al. (2021, 2023) are shown in black diamonds, representing excellent coverage of the space parameters. The diamond symbols show the whole coverage of the libraries—not only the models selected to match the observed stars. The bottom-right panel shows the histograms of distances (in magnitudes) between the star and its associated model. The histogram is normalized by the number of stars in each evolutionary phase, where the vertical axis represents the percentage of stars in that specific evolutionary stage. We find that for all evolutionary phases, the distance distributions show a narrow profile peaked around 0.2 mag, as shown in the last panel of Figure 3.

Figure 4 shows the same as Figure 3 for M 2, using the F. Castelli & R. L. Kurucz (2003) library that matches the low metallicity of this cluster. The peaks of the distance distributions for this case are around 0.1 mag, which is less than that of the NGC 2808 for all the evolutionary phases. This is a consequence of the F. Castelli & R. L. Kurucz (2003) library having higher sampling densities and temperatures in the $T_{eff}$–$\log g$ parameter space, thus achieving bluer colors.

### 6. Synthetic Integrated Spectra

Once each star in the CMD is matched to a stellar spectrum, the modeled integrated flux $F_{SSP}(\lambda)$ of the GC is given by the sum of weighted contributions from all stars along the spectrum (Equation (1)):

$$F_{SSP}(\lambda) = \sum_{i=1}^{N} f_{star}(\lambda) \times 10^{-0.4(M_i - M_s)}, \quad (1)$$

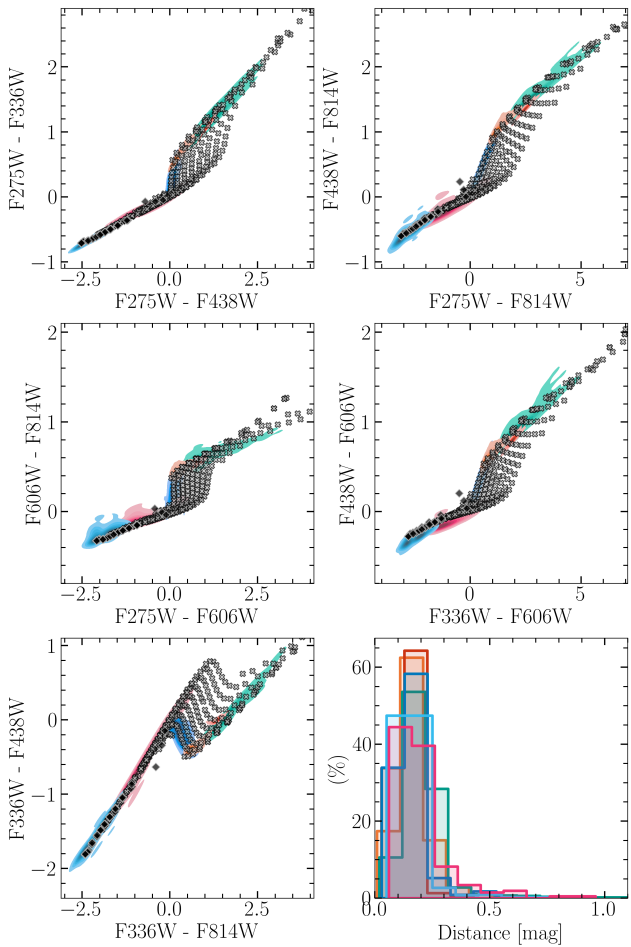

**Figure 3.** Color–color diagrams for NGC 2808, with stars hued by evolutionary phase: MS in orange, GB in teal, RHB in red, BS in blue, BHB in magenta, and EHB in cyan. These indicate that each evolutionary phase occupies a distinct locus in multiband color space. Bottom right: histogram showing the distribution of distances between observed and synthetic colors, normalized by the number of stars in each evolutionary phase. This quantifies the tightness of the match. The color scheme follows that of the other panels.

where $f_{star}(\lambda)$ is the synthetic spectrum that represents the $i$th star and N is the total number of stars in the CMD (R. P. Schiavon et al. 2004a; L. P. Martins et al. 2019). Each synthetic stellar spectrum is weighted by the flux in the F438W filter, chosen as an intermediate wavelength of the photometric data. The weighting factor is obtained based on the differences between the model magnitude, $M_i$, and its corresponding observed stellar magnitude, $M_s$, and it accounts for the dilution factor $R/d$, where $R$ is the radius of the star and $d$ is its distance (L. Casagrande & D. A. VandenBerg 2014). We integrated the spectra for six subsamples, one per evolutionary phase, as described in Section 5.

The integrated-light spectra of MS, GB, RHB, BS, BHB, and EHB are shown in Figure 5 for NGC 2808 in orange, teal, red, blue, cyan, and magenta, respectively. The resolution in this case is limited to the resolving power of the P. R. T. Coelho (2014) grid—that is, $R \approx 20{,}000$, ranging from 2000 to 9 000 Å.

Figure 6 shows the integrated-light spectra for the six subsamples of evolutionary phases of M 2. Compared to

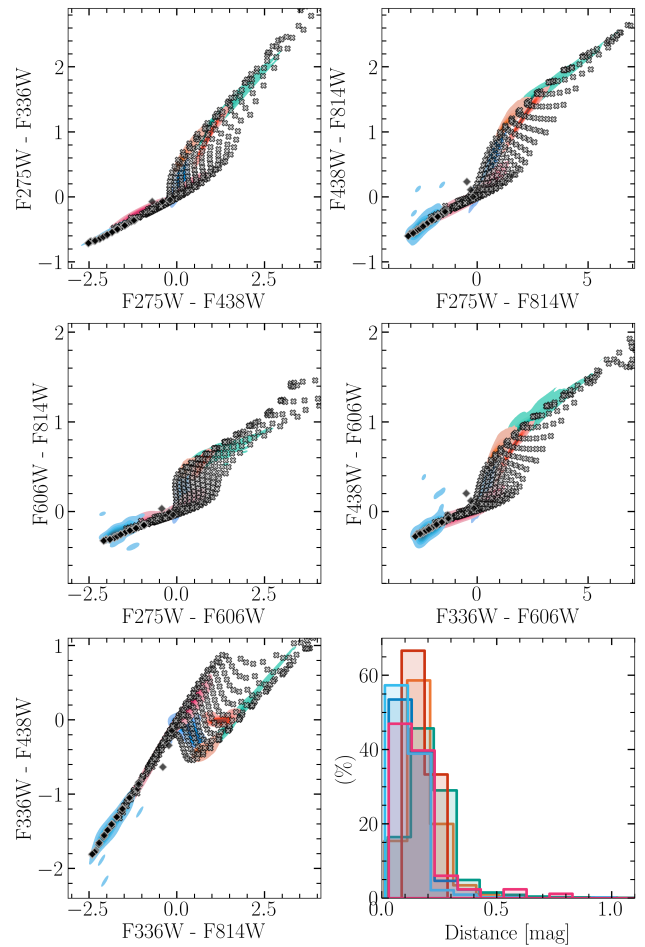

**Figure 4.** Color–color diagrams and histogram of the distribution of normalized distances between observed and synthetic colors for M 2, with stars hued by evolutionary phase following the same scheme as in Figure 3.

Figure 5, it follows the same color scheme and covers a wider wavelength range that extends into the far-UV, from 1000 Å. The resolution power in this case is limited to the F. Castelli & R. L. Kurucz (2003) grid ($R \approx 250$).

In both GCs, the light from MS and GB dominates in the redder range, spanning both visible and IR wavelengths. However, the blue and extreme HBs are dominating in the bluer range, thus affecting the UV part of the SSP. This illustrates the impact of different evolutionary phases, revealing the diversity in the overall integrated light of a stellar population.

## 7. SSP Analysis

To quantify the contribution of each hot stellar component within the SSP, we constructed synthetic populations by combining different sets of evolutionary phases. We define a reference spectrum, hereafter "base," composed of MS, GB, and RHB stars. Additional SSPs were generated by incrementally adding one hot component, BS, BHB, or EHB, to the base model. We also compute a fifth SSP that includes all evolutionary phases, hereafter "all." By comparing the relative fluxes of these five synthetic SSPs with those of the base model, we evaluated the contribution of each hot component to the integrated spectrum.

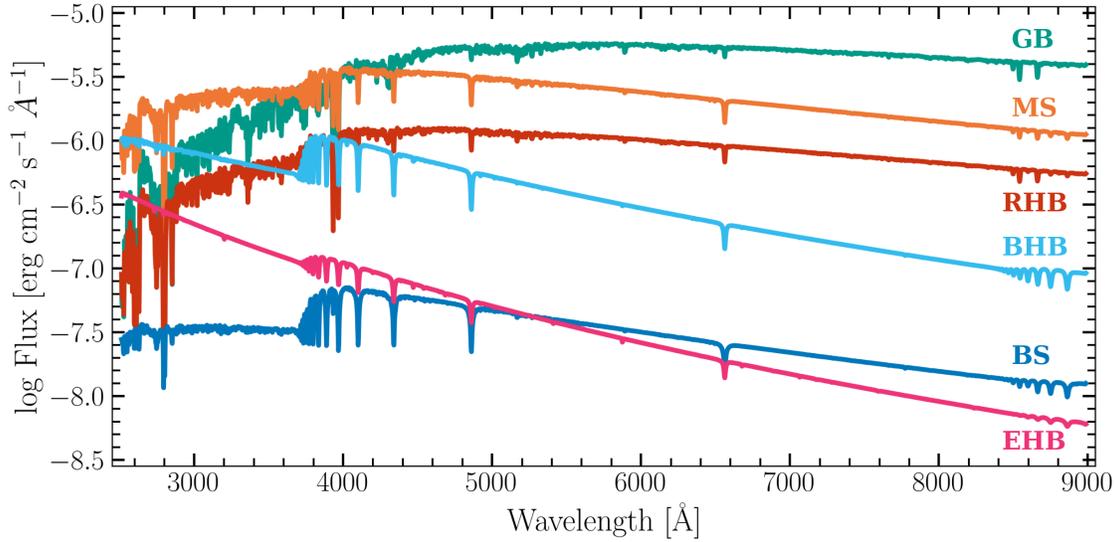

**Figure 5.** Integrated spectra for each evolutionary phase of NGC 2808, color-coded as follows: MS (orange), GB (teal), RHB (red), BS (blue), BHB (cyan), and EHB (magenta). Note that BHB and EHB are as important as MS and GB in the blue wavelengths (below 3000 Å).

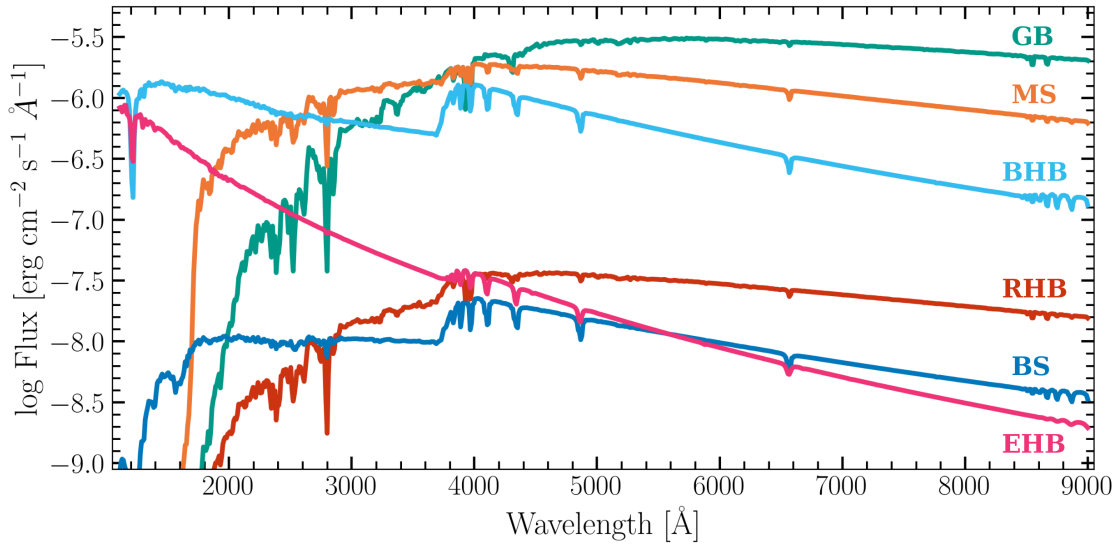

**Figure 6.** Integrated spectra for each evolutionary phase of M 2, color-coded as follows: MS (orange), GB (teal), RHB (red), BS (blue), BHB (cyan), and EHB (magenta). Note that this plot covers bluer wavelengths than Figure 5, and in this case, BHB and EHB dominate the UV flux (below 1500 Å).

### 7.1. Impact on Integrated Fluxes

*Models based on NGC 2808.* In Figure 7, we present the five SSP spectra constructed from the CMD of NGC 2808, degraded to a resolution of $R \approx 1000$ for visualization. The solid gray line represents the base model, while additional hot components are overplotted: the BHB in cyan dashed lines, BS in blue dotted–dashed lines, and the EHB in magenta dotted lines. Among the hot phases, the BHB dominates the UV contribution, accounting for 16.9% of the total SSP flux. The EHB also contributes significantly, at 3.3%, while the BS component adds a modest 0.9%. When all hot components are included, the integrated flux in the near-UV ($\lambda \approx 3200$ Å) increases by approximately 21% compared to the base model. This result underscores the importance of including hot, low-mass evolutionary phases in SSP modeling to accurately reproduce the integrated light of old stellar populations.

*Models based on M 2.* In Figure 8, we present the five SSPs constructed from the M 2 CMD using the same color scheme as in Figure 7. The general trends are similar to those found for NGC 2808, but here the models extend further into UV. As noted in Section 6, the resolving power of the M 2 spectra is approximately $R \approx 250$. In the far-UV region, the EHB flux becomes comparable to that of the BHB, reinforcing the importance of including hot stellar components in population synthesis models to properly account for the integrated UV light of old stellar systems.

### 7.2. Age Impact

An important aspect of this work is to evaluate how the inclusion of different stellar components affects ages derived by inverse methods, such as SED and spectral fitting. To do so, we used the five synthetic spectra of NGC 2808 described in the first paragraph of Section 7 as proxies for stellar populations, mimicking GCs with and without hot-evolved stellar components.

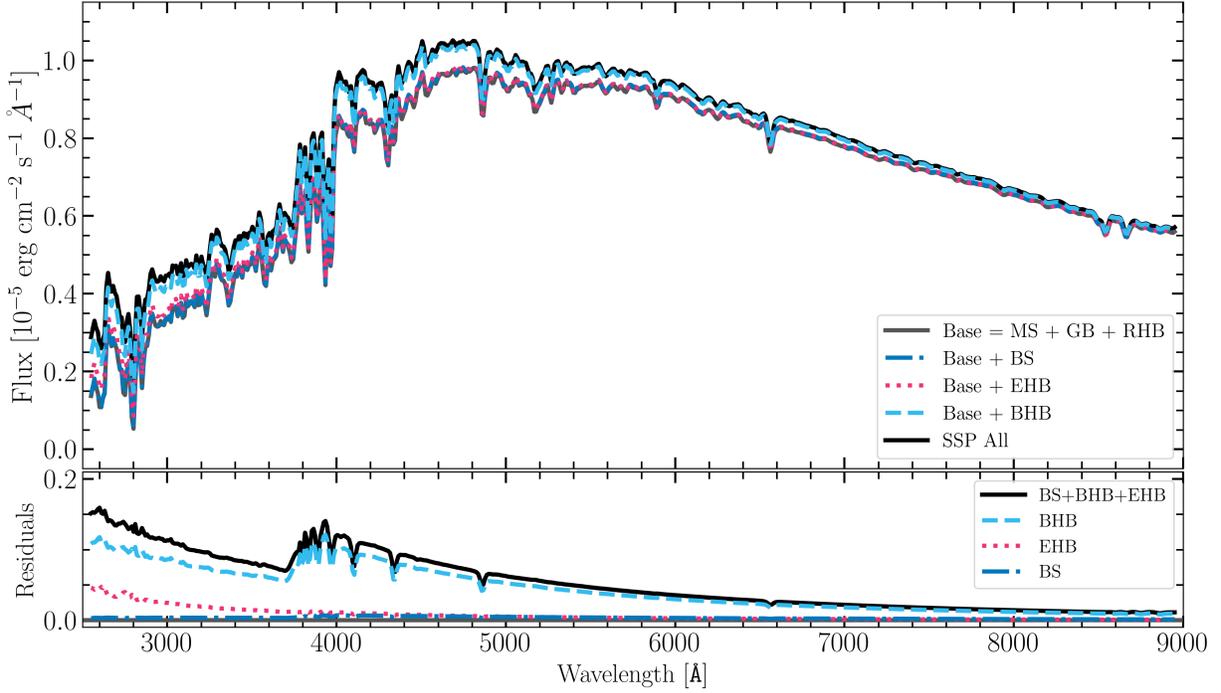

**Figure 7.** SSP spectra for NGC 2808, including the base model (gray solid line) and models with additional contributions from the BS (blue dotted–dashed line) that closely match the base model, BHB (cyan dashed), and EHB (magenta dotted) stars. The final composite spectrum, including all components, is also shown as a black solid line. The bottom panel displays the residuals with respect to the base model ($F_i - F_{\rm base}$), highlighting the relative flux contributions of each hot stellar component.

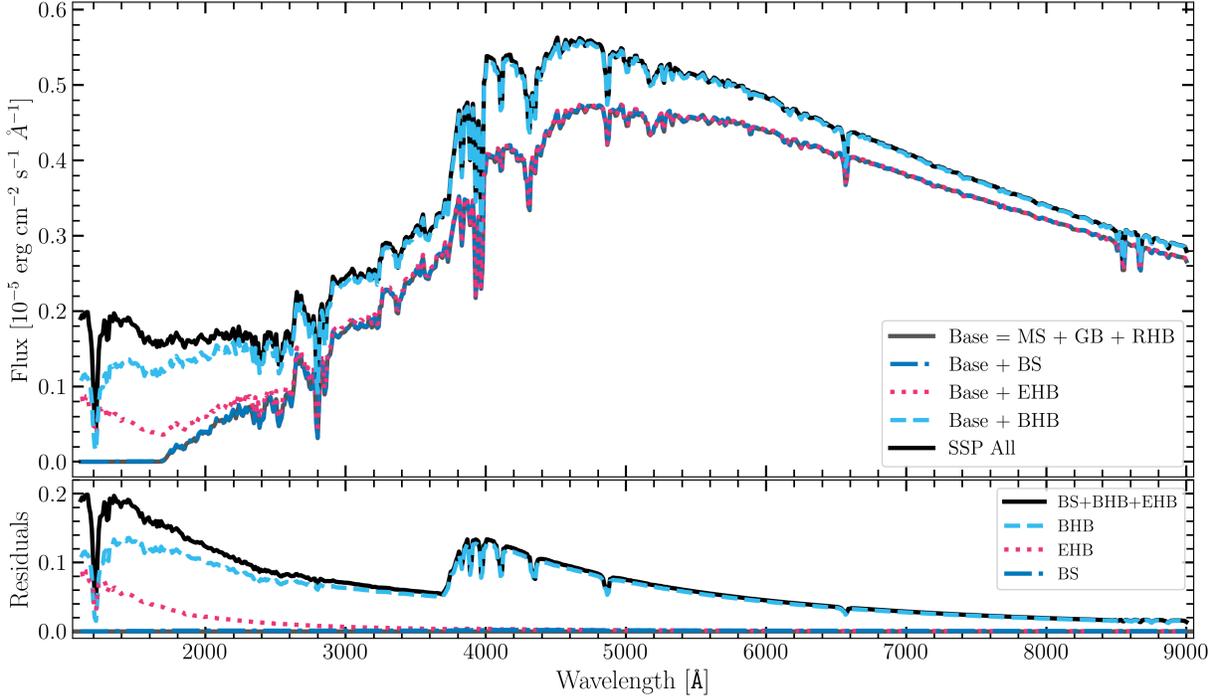

**Figure 8.** SSP spectra for M 2, including different hot stellar components and residuals ($F_i - F_{\rm base}$), following the same color scheme as in Figure 7. While the BS still have a small contribution around 2000 Å, the extended wavelength coverage into the far-UV allows a direct comparison of the relative flux contributions from BHB and EHB stars.

We simulated age estimates for the synthetic GCs using the spectral fitting code Starlight (R. Cid Fernandes et al. 2005), with the BC03 library (G. Bruzual & S. Charlot 2003) as the stellar population base. The SSP models assume an IMF of G. Chabrier (2003) and Padova isochrones (A. Bressan et al. 2012). Table 3 presents the resulting ages, including a comparison with the value derived from the observed integrated spectrum of NGC 2808 (R. P. Schiavon et al. 2005), estimated to be ∼8.8 Gyr. This is younger than the ∼11.5 Gyr age inferred from the spectroscopy of individual stars.

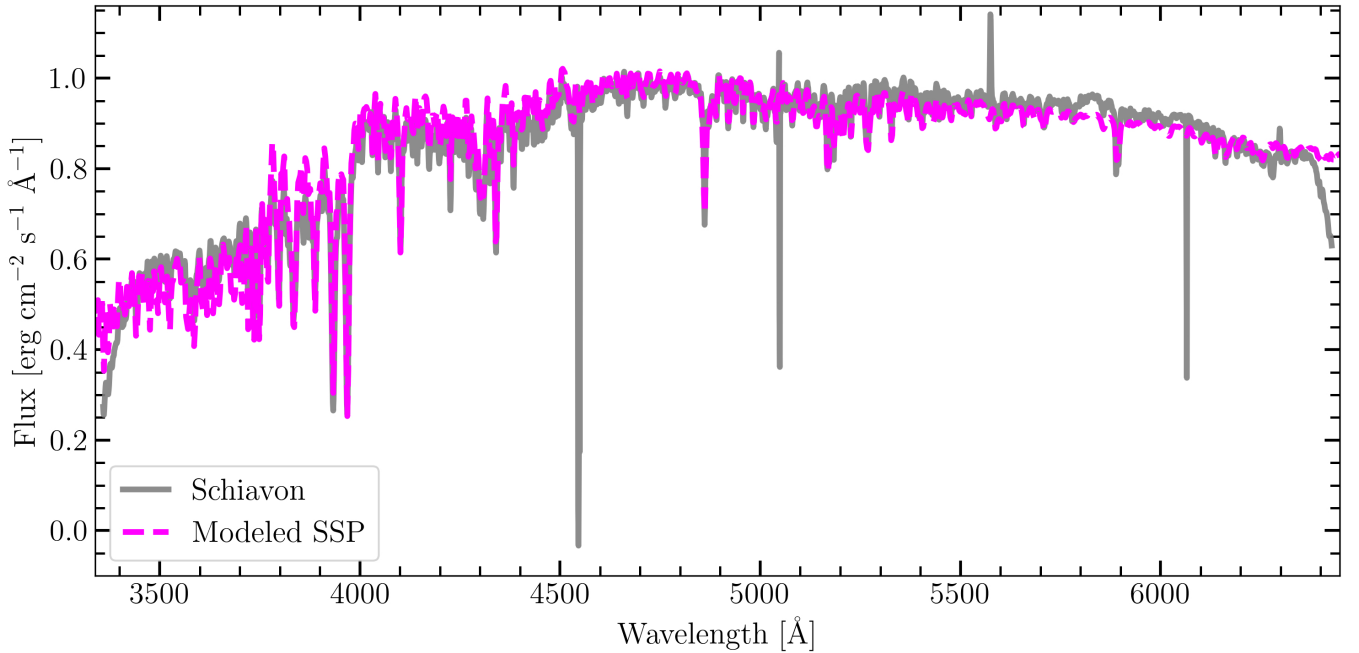

**Figure 9.** Synthetic all SSP in magenta compared to the observed integrated spectrum of NGC 2808 from the Blanco Telescope (R. P. Schiavon et al. 2005), shown in gray, covering the optical region and normalized to 1 at wavelength 4995 Å. The drops in observed flux at the extremes are due to instrumental effects.

**Table 3**
Ages for NGC 2808 Derived from BC03 Models for Different Synthetic Populations and the Observed Spectrum

| Sample | Age$_{median}$ (Gyr) |
| --- | --- |
| Base | 10.18 |
| Base + BHB | 8.49 |
| Base + EHB | 10.16 |
| Base + BS | 10.33 |
| All model | 4.98 |
| Observed | 8.81 |

The origin of such a discrepancy is identified as the contribution of blue light from hot low-mass stars being misinterpreted as emission from young stars by Starlight. This is expected, as the BC03 library does not include evolved hot stellar types in its synthesis. In fact, the base model, which excludes hot stellar components, yields an age of more than 10 Gyr, which is substantially closer to the isochrone age of NGC 2808.

Adding BHB stars to the base model reduces the estimated age to ∼8.5 Gyr, approaching the result derived from the observed spectrum (∼8.8 Gyr). This supports the idea that blue light from BHB stars may be misidentified as a young stellar component in spectral fitting. In contrast, the inclusion of EHB or BS components alone has a minor effect on the age derived from optical spectra, as their contribution is modest compared to that of BHB.

However, when all evolved hot stellar phases (BHB, EHB, and BS) are included in the base model, the estimated age decreases to ∼5 Gyr, illustrating how strongly blue-evolved stars can bias age determinations if not properly accounted for. This highlights the high sensitivity of integrated-light age estimates to the presence of evolved hot, low-mass stellar components.

These results reinforce the importance of incorporating evolved hot stars, particularly HB stars, into SSP models. Accurately modeling these populations is crucial for disentangling the effects of age and HB morphology in the spectral analysis of unresolved stellar systems.

### 7.3. Comparison with Observations

Integrated light spectra of the Galactic GCs NGC 2808 and M 2 are available in the literature (R. P. Schiavon et al. 2005). However, the field of view (FoV) at the Blanco Telescope, which was used to obtain these spectra, differs significantly from that of the HST photometry employed to tag stars to evolutionary phases in the CMD. Consequently, the modeled spectrum is not expected to match the observed spectrum with exact correspondence because of the GC sampling difference, possibly presenting radial variations of the population in the cluster. Nevertheless, the comparison provides a valuable consistency check for the reliability of our stellar population synthesis models.

Figure 9 compares our synthetic all SSP (magenta), which includes all evolutionary phases, to the integrated spectrum (gray) of NGC 2808 from R. P. Schiavon et al. (2005). Both spectra were normalized to 1 at wavelength 4995 Å. The overall agreement is good across the optical range, including the bluer wavelengths, where the contribution from evolved hot stellar populations is strongest. This agreement supports the physical plausibility of our models and their ability to reproduce integrated spectra with evolved hot stars.

Residuals between the observed spectrum and both the base and all SSP models and the BC03 SSP with 11.5 Gyr shown in

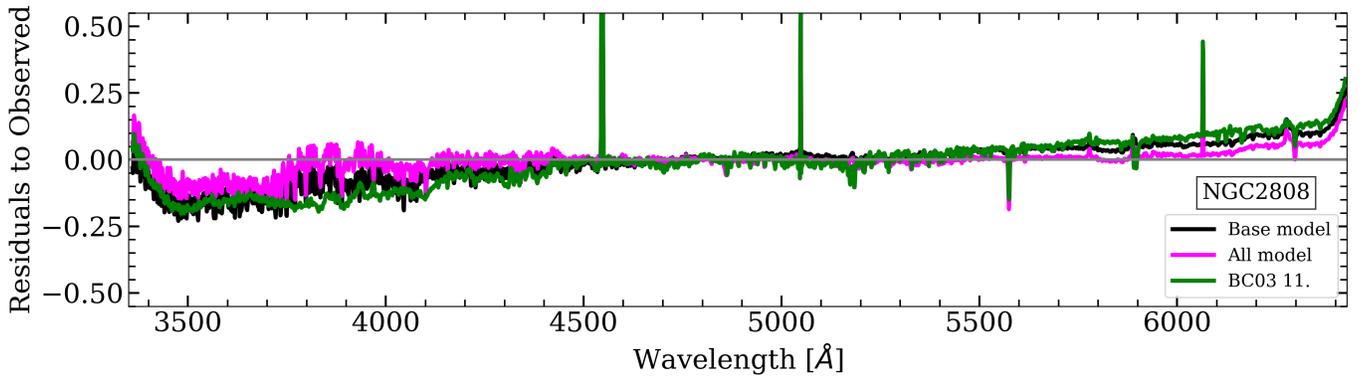

**Figure 10.** Residuals of the synthetic base SSP (in black), all SSP (in magenta), and 11.5 Gyr BC03 SSP (in green) with respect to the observed integrated spectrum of NGC 2808 from the Blanco Telescope (R. P. Schiavon et al. 2005), represented by the gray zero line, spanning the optical region. The residuals are lower for our all SSP.

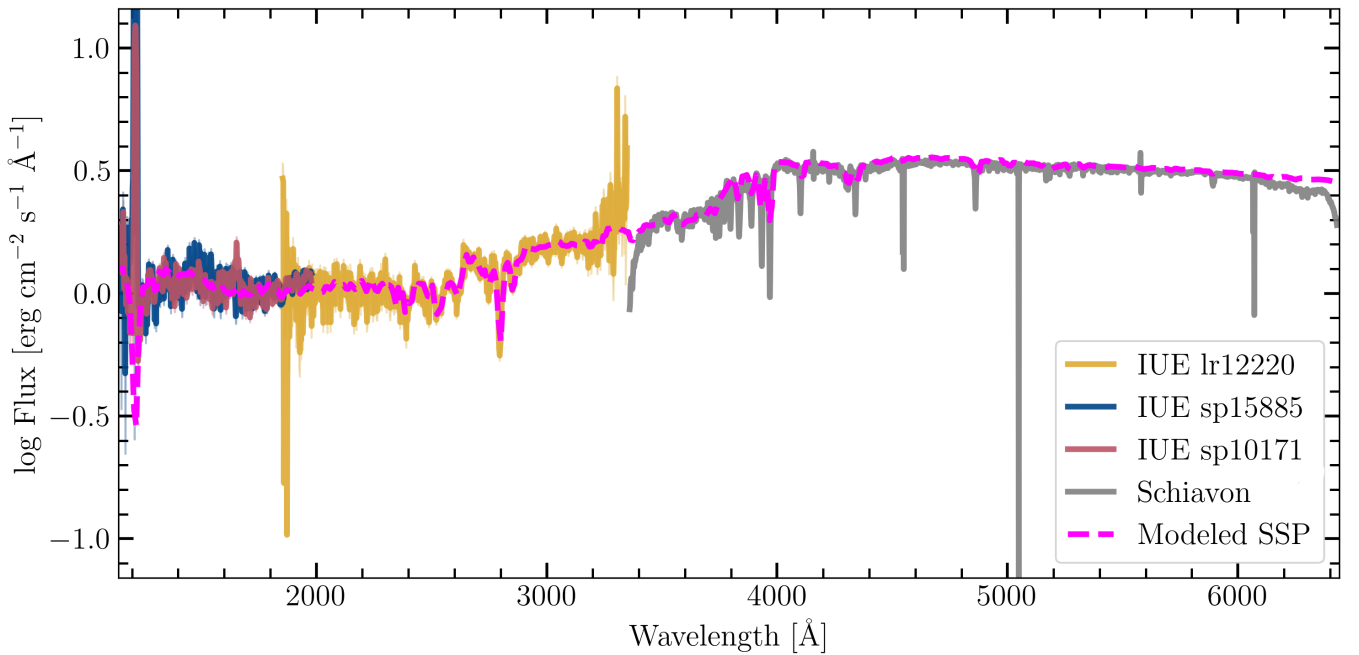

**Figure 11.** Synthetic all SSP in magenta compared to the observed integrated spectra of M 2 from IUE (V. Caloi et al. 1983) in blue, red, and yellow, covering the far-UV normalized to 2750 Å, and Blanco Telescope (R. P. Schiavon et al. 2005) in gray, covering the optical normalized to 4995 Å.

Figure 10 emphasize the importance of including hot, low-mass stellar components in the modeling of integrated light of GCs. The residual differences in flux (model minus observed, in units of erg s$^{-1}$ cm$^{-2}$ Å$^{-1}$) are systematically lower when all hot evolutionary phases (BHB, EHB, and BS) are included, particularly at shorter wavelengths. This shows that incorporating these components improves the agreement between synthetic and observed spectra, highlighting their critical role in accurate stellar population modeling.

A similar comparison of our synthetic all SSP (magenta) and observed spectra (blue, red, yellow, and gray) is shown for M 2 in Figure 11. In this case, observations span a broader wavelength range with data from far-UV (IUE; V. Caloi et al. 1983) to optical (Blanco Telescope; R. P. Schiavon et al. 2005). The three IUE spectra were normalized to wavelength 2750 Å, and the Blanco spectrum was normalized to 4995 Å. Our synthetic all SSP spectrum reproduces the global shape and flux levels across this extended spectral range remarkably well, making this the first time that an integrated-light model of a GC has been constructed with this level of completeness and accuracy, from the far-UV through the optical.

Figure 12 also highlights the essential role of UV observations in revealing the contributions of evolved hot stars, especially in GCs such as M 2 that host extended HB populations. The residuals shown in Figure 12 demonstrate that models that omit these evolved hot phases underestimate the flux in the UV and blue optical regions. This analysis, including UV data, suggests that the comparison between local and redshifted population ages may be prone to biases if hot-evolved components are not properly considered.

As discussed in Section 3 and Appendix B, completeness declines at the faint end of the MS, leading to a slight underrepresentation of cool stars. However, since these stars contribute little to the total flux compared to other evolutionary phases, the effect on the integrated spectra is negligible. Also, the residual flux calibration artifacts in the observed spectrum may contribute to the small red residuals seen in Figures 10 and 12.

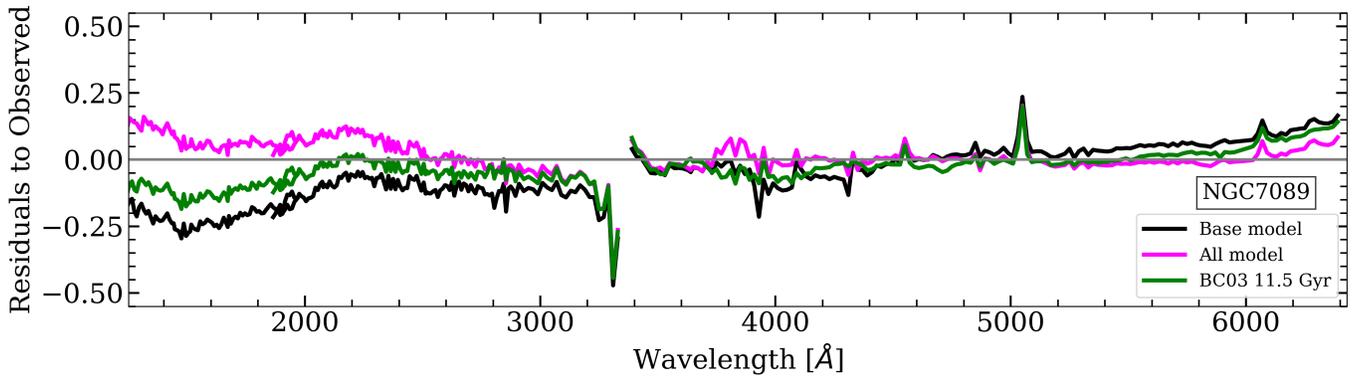

**Figure 12.** Residuals of the synthetic base SSP (in black), all SSP (in magenta), and 11.5 Gyr BC03 SSP (in green) compared to the observed integrated spectrum of M 2 from IUE (V. Caloi et al. 1983) and Blanco Telescope (R. P. Schiavon et al. 2005), represented by the gray zero line, ranging from the far-UV to the optical regions. Much like in Figure 10, our all SSP presents an overall lower residual compared to the others.

The results for these well-known GCs demonstrate that spectral population synthesis models that neglect evolved hot stars, such as the BHB, EHB, and BS components, introduce systematic biases in age estimates derived from integrated light. The blue flux from these phases can mimic the signature of genuinely young populations, leading to an underestimation of the ages of GCs. This bias is particularly relevant for studies that rely on the u-band, such as those using LSST (Ž. Ivezić et al. 2019), J-PLUS (A. J. Cenarro et al. 2019; C. López-Sanjuan et al. 2021), or J-PAS (N. Benitez et al. 2014) photometric catalogs. As these surveys extend to faint magnitudes and large volumes, properly accounting for hot-evolved populations becomes essential for age and metallicity estimates in extragalactic stellar populations. For example, analyses of stellar halos or galaxies at intermediate redshifts that use UV-optical colors or SEDs can be significantly affected. Accurately modeling the contribution from hot subdwarfs and other evolved stars is therefore crucial to avoid misinterpreting composite populations. Our analyses confirm the perspective that has been claimed in the literature, based on various experiments (see Section 1). Including these stellar phases improves the fidelity of spectral fits, particularly at shorter wavelengths. We expect that with our proposed SSP modeling, the ages inferred from inverse methods will yield better age estimates, in better agreement with those obtained from resolved stellar populations.

### 8. Summary and Conclusions

We modeled the integrated light from different stellar evolutionary phases in Galactic GCs using high-quality CMDs (D. Nardiello et al. 2018) and state-of-the-art synthetic stellar libraries that include hot, high-gravity, low-mass stars (T. A. Pacheco et al. 2021, 2023). Our novel CMD-based approach, which matches evolutionary phases in a 10-dimensional color space, allows us to build SSPs with realistic HB morphologies and quantify the spectral contribution of each evolutionary phase to the integrated spectrum. In far-UV (wavelengths ≲1700 Å), evolved hot stars can account for up to 100% of the flux, indicating the dominant role of BHB and EHB stars in this range. Though nearly invisible in the optical, these stars are dominant UV contributors and play a key role in modeling the spectra of unresolved stellar populations.

To assess the impact on age estimates, we compared results from the Starlight spectral fitting code (R. Cid Fernandes et al. 2005), using BC03 templates (G. Bruzual & S. Charlot 2003), against both synthetic and observed spectra of NGC 2808. We found that models including BHB, EHB, and BS stars consistently yield younger ages, thus indicating that blue flux from hot, evolved stars can be misinterpreted as evidence of recent star formation, a bias clearly reflected in the residuals and best-fit ages of both synthetic and observed spectra.

By explicitly incorporating these hot-evolved phases, our SSP models significantly reduce residuals in spectral fitting, especially in the UV and blue ranges, leading to more accurate and physically consistent age estimates. This improvement contributes to resolving the well-known challenge of age–HB morphology degeneracy that complicates the interpretation of integrated light from old stellar populations.

Although our results lead to improved age estimates, a complete understanding of the physical processes governing GC formation and evolution remains elusive. Further investigation is required to enhance stellar population synthesis models, particularly for GCs with extended HBs. Our findings have broader implications for tracing galaxy assembly, as accurately dating unresolved stellar populations is crucial for reconstructing the SFH and chemical enrichment of galaxies.

Our study highlights the importance of developing comprehensive SSP libraries that incorporate the full range of stellar evolutionary phases. Such models are essential for robust population synthesis and for reliably interpreting the integrated light of both local and distant (high redshift) systems.

### Acknowledgments

This study was financed in part by the Coordenação de Aperfeiçoamento de Pessoal de Nível Superior—Brasil (CAPES) —Finance Code 001. T.A.P. acknowledges funding from the Santander International Mobility Program (PRPG–39/2022) and support by Conselho Nacional de Desenvolvimento Científico e Tecnológico (CNPq) under the doctoral stay program (200491/ 2022-9). T.A.P. appreciates the hospitality of the Astrophysics Research Institute at Liverpool John Moores University for a visit during which some of the work reported in this paper was developed. T.A.P., A.L.C.S., and C.J.B. acknowledge funding from Fundação de Amparo à Pesquisa do Rio Grande do Sul (FAPERGS)—23/2551-0001832-2. P.R.T.C. acknowledges support from CNPq (310555/2021-3) and Fundação de Amparo à Pesquisa do Estado de São Paulo (FAPESP; 2021/08813-7). L.P. M. thanks FAPESP–2022/03703-1 and CNPq–307115/2021-6. E.V.R.L. acknowledges the financial support given by CAPES–88887.470064/2019–00, CNPq–169181/2017-0, and FAPESP–2024/15229-8. M.P.D. thanks CNPq for support under grant CNPq–305033. A.L.C.S. acknowledges funding from CNPq–314301/2021-6 and 445231/2024-6. R.R. acknowledges support from CNPq–445231/2024-6, 311223/2020-6, 404238/2021-1, and 310413/2025-7; FAPERGS–19/1750-2 and 24/2551-0001282-6; and CAPES–88881.109987/2025-01. This research has made use of the Spanish Virtual Observatory (SVO) Filter Profile Service "Carlos Rodrigo," funded by the Spanish Ministry of Science and Innovation/Agencia Estatal de Investigación grant MCIN/AEI/10.13039/501100011033/ through grant PID2023-146210NB-I00.

## Appendix A
## CMD Cutoffs

Tables 4 and 5 present a comprehensive range of photometric cutoffs in various magnitudes and colors to display the subsamples for NGC 2808 and M 2, respectively.

**Table 4**
Cutoffs (in mag) of the NGC 2808's Evolutionary Phase Subsamples by Using the TOPCAT

|  | MS | GB | RHB | BS | BHB | EHB |
|---|---|---|---|---|---|---|
| F275W | $\geqslant 18.0$ | ⋯ | $\leqslant 17.8$ | $\geqslant 18.6$ | $\leqslant 17.0$ | ⋯ |
| F336W | $\geqslant 17.5$ | $\leqslant 18.8$ | $\leqslant 15.9$ | $\geqslant 18.6$ | $\leqslant 17.3$ | ⋯ |
| F606W | ⋯ | ⋯ | 15.0 to 15.8 | 15.8 to 18.8 | $\leqslant 18.0$ | ⋯ |
| F814W | $\geqslant 17.15$ | $\leqslant 17.5$ | ⋯ | ⋯ | ⋯ | ⋯ |
| $C_{F(275-336)W}$ | 0.0 to 2.0 | $\geqslant 0.75$ | $\geqslant 0.5$ | −0.5 to 0.9 | −0.4 to 0.5 | $\leqslant -0.4$ |
| $C_{F(336-438)W}$ | −1.0 to 1.0 | $\geqslant -0.2$ | ⋯ | −0.7 to 0.0 | ⋯ | ⋯ |
| $C_{F(438-606)W}$ | 0.0 to 2.0 | $\geqslant -0.7$ | $\leqslant 1.0$ | 0.0 to 0.75 | $\leqslant 0.4$ | ⋯ |
| $C_{F(606-814)W}$ | 0.0 to 1.3 | $\geqslant 0.35$ | 0.3 to 0.7 | −0.05 to 0.35 | $\leqslant 0.3$ | $\leqslant 0.0$ |
| $C_{F(336-814)W}$ | ⋯ | ⋯ | ⋯ | ⋯ | ⋯ | $\leqslant 0.35$ |
| Notes | no GB | ⋯ | ⋯ | ⋯ | ⋯ | ⋯ |

**Table 5**
Cutoffs (in mag) of the M 2's Evolutionary Phase Subsamples by Using the TOPCAT

|  | MS | GB | RHB | BS | BHB | EHB |
|---|---|---|---|---|---|---|
| F275W | $\geqslant 18.4$ | ⋯ | $\leqslant 18.0$ | $\geqslant 18.69.0$ | $\leqslant 17.2$ | ⋯ |
| F336W | $\geqslant 17.9$ | $\leqslant 18.95$ | $\leqslant 16.5$ | $\geqslant 18.9$ | $\leqslant 17.5$ | ⋯ |
| F606W | ⋯ | ⋯ | 15.0 to 15.8 | 16.2 to 19.2 | $\leqslant 18.2$ | ⋯ |
| F814W | $\geqslant 17.45$ | $\leqslant 17.65$ | ⋯ | ⋯ | ⋯ | ⋯ |
| $C_{F(275-336)W}$ | 0.15 to 2.15 | $\geqslant 0.8$ | $\geqslant 0.5$ | −0.5 to 0.9 | −0.4 to 0.5 | $\leqslant -0.4$ |
| $C_{F(336-438)W}$ | −0.85 to 1.85 | $\geqslant -0.25$ | ⋯ | −0.5 to 0.2 | ⋯ | ⋯ |
| $C_{F(438-606)W}$ | 0.15 to 2.15 | $\geqslant -0.75$ | $\leqslant 1.05$ | 0.0 to 0.75 | $\leqslant 1.05$ | ⋯ |
| $C_{F(606-814)W}$ | 0.15 to 1.45 | $\geqslant 0.4$ | 0.4 to 0.75 | −0.05 to 0.4 | $\leqslant 0.4$ | $\leqslant 0.0$ |
| $C_{F(336-814)W}$ | ⋯ | ⋯ | ⋯ | ⋯ | ⋯ | $\leqslant 0.35$ |
| Notes | no GB | ⋯ | ⋯ | ⋯ | ⋯ | ⋯ |

## Appendix B
## Completeness

To assess the photometric completeness under our selection criteria, we analyzed artificial stars following the same error thresholds applied to the real data (see Section 3). A star was considered successfully recovered if its input and output positions differed by less than 0.5 pixels and its recovered magnitude by less than 0.75 mag. This analysis, presented in Figure 13, confirms that the dominant contributors to the integrated light remain well sampled.

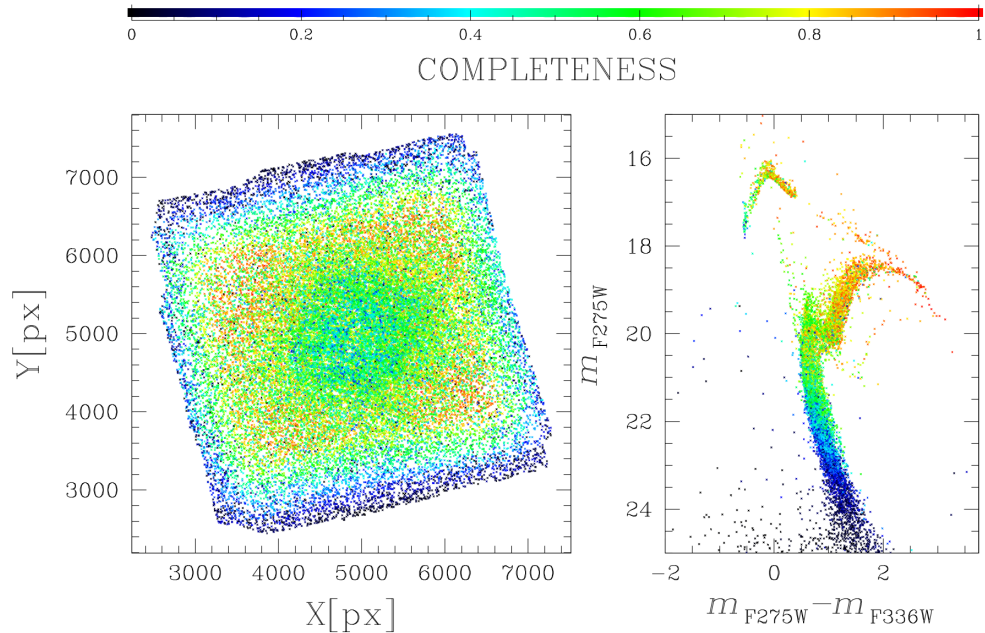

**Figure 13.** Photometric completeness after applying our selection criteria in the NGC 7089 HST data, hued by the rainbow (where red represents 100% complete) for positions on the left and UV CMD for positions on the right.

## ORCID iDs

Thayse A. Pacheco ● https://orcid.org/0000-0002-8139-7278
Paula R. T. Coelho ● https://orcid.org/0000-0003-1846-4826
Lucimara P. Martins ● https://orcid.org/0000-0002-3666-2810
Ricardo P. Schiavon ● https://orcid.org/0000-0002-2244-0897
Erik V. R. de Lima ● https://orcid.org/0000-0002-6268-8600
Marcos P. Diaz ● https://orcid.org/0000-0002-6040-0458
Domenico Nardiello ● https://orcid.org/0000-0003-1149-3659
Ronaldo S. Levenhagen ● https://orcid.org/0000-0003-2499-9325
Rogério Riffel ● https://orcid.org/0000-0002-1321-1320
Charles J. Bonatto ● https://orcid.org/0000-0002-4102-1751
Ana L. Chies-Santos ● https://orcid.org/0000-0003-3220-0165